\documentclass[preprint]{elsarticle}

\usepackage{hyperref,subcaption,tabularx}
\usepackage[margin=2.5cm]{geometry}
\graphicspath{{./images/}}
\usepackage[utf8]{inputenc}
\usepackage[T1]{fontenc}
\usepackage{parskip}
\usepackage{xcolor}

\journal{Acta Materialia}

\begin{document}
\begin{frontmatter}

\title{Low temperature deformation of MoSi$_2$ and the effect of Ta, Nb and Al as alloying elements}

\author[1]{Carolin Zenk}

\author[2]{James S.K.-L. Gibson}

\author[3]{Verena Maier-Kiener}

\author[1]{Steffen Neumeier}

\author[1]{Mathias Göken}

\author[2]{Sandra Korte-Kerzel\corref{corr1}}
\ead{korte-kerzel@imm.rwth-aachen.de}

\address[1]{Friedrich-Alexander-Universität Erlangen-Nürnberg, Materials Science \& Engineering I, Martensstr. 5, D-91058 Erlangen, Germany}
\address[2]{RWTH Aachen University, Institute of Physical Metallurgy and Materials Physics, Kopernikusstr. 14, D-52074 Aachen, Germany
}
\address[3]{Montanuniversität Leoben, Department Physical Metallurgy and Materials Testing, Roseggerstr. 12, A-8700 Leoben, Austria}

\begin{abstract}
Molybdenum disilicide (MoSi$_2$) is a very promising material for high temperature structural applications due to its high melting point (2030$^{\circ}$C), low density, high thermal conductivity and good oxidation resistance. However, MoSi$_2$ has limited ductility below 900$^{\circ}$C due to its anisotropic plastic deformation behaviour and high critical resolved shear stresses on particular slip systems.

Nanoindentation of MoSi$_2$ microalloyed with aluminium, niobium or tantalum showed that all alloying elements cause a decrease in hardness. Analysis of surface slip lines indicated the activation of the additional \{1 1 0\}$<$1 1 1$>$ slip system in microalloyed MoSi$_2$, which is not active below 300$^{\circ}$C in pure MoSi$_2$. This was confirmed by TEM dislocation analysis of the indentation plastic zone. Further micropillar compression experiments comparing pure MoSi$_2$ and the Ta-alloyed sample enabled the determination of the critical resolved shear stresses of individual slip systems even in the most brittle [0~0~1] crystal direction.
\end{abstract}


\end{frontmatter}

\section{Introduction}
The improvement of high temperature, structural materials plays a particularly significant role in the development of power-plant turbine engines, where application temperatures and efficiency are directly linked. Nickel-base superalloys are currently used for components under the highest thermo-mechanical loads, but their application is limited to temperatures below 1150$^{\circ}$C. Molybdenum disilicides (MoSi$_2$) and MoSi$_2$-composites are promising candidate materials to further increase operational temperatures. They can be used in an oxidising atmosphere up to 1500$^{\circ}$C, and additionally have a low density, good corrosion resistance, good thermal conductivity and excellent oxidation behaviour at temperatures above 1000$^{\circ}$C as a result of a continuous protective SiO$_2$ oxide layer~\cite{1,2,3,4,5,ISI:000348048800131,LIANG2018389}. However, current drawbacks include the oxidation resistance at intermediate temperatures~\cite{5,6}, and, for rotational components, the ability to form a strong protective surface layer, as well as the overall mechanical properties. The lack of room temperature ductility and as well as high temperature strength therefore currently limits the application of molybdenum disilicides to areas with lower mechanical stresses like heating elements~\cite{1,2,3,7}. Some recent work in the literature is beginning to bridge this gap, by using MoSi$_2$ as part of thermal barrier coatings~\cite{ISI:000398483700018,ISI:000381143200001,ISI:000402743300011} where these materials are in less mechanically-critical demands, but are nevertheless still exposed to rotational forces.

One approach to improve ductility is microalloying MoSi$_2$ with aluminium, niobium, tantalum, magnesium, chromium, zirconium or vanadium~\cite{8,9,10,11,12,13,14}. The influence of these alloying elements can be estimated by calculation of the stacking fault energies of individual slip systems~\cite{15}. However - especially in the low temperature regime - the experimental verification is limited to the study of the dissociation width of dislocations and stacking faults. Basic mechanical properties such as the critical resolved shear stress (CRSS) of particular slip systems cannot be analysed for brittle materials like MoSi$_2$ using conventional methods. The underlying plastic deformation anisotropy hinders the extrapolation of hardness and therefore single crystal experiments are needed, but macroscopic uniaxial compression of MoSi$_2$ commonly results in brittle fracture instead of significant plastic deformation. These experimental challenges can be overcome by nanomechanical testing~\cite{16,17}, which allows direct comparison between pure MoSi$_2$ and the more ductile microalloyed disilicides, including quantitative measurement of the CRSS.

In this work, nanoindentation has been used to study the plastic deformation behaviour of MoSi$_2$ microalloyed with aluminium, niobium or tantalum. Slip lines that arose on the surface after indentation were studied and correlated to statistically activated slip planes. Further studies using transmission electron microscopy (TEM) and micropillar compression tests were carried out on (Mo,Ta)Si$_2$, as little is known about its effect as an alloying element. TEM analysis of the dislocation structures directly underneath indents was performed in MoSi$_2$ and (Mo,Ta)Si$_2$. Micropillar compression experiments were conducted to measure CRSS values, where the uniaxial stress state and the suppression of brittle fracture allow an orientation-dependent analysis of plasticity~\cite{18}. (For a review on the mechanisms leading to the prevalence of plasticity in such small samples and other size effects observed in microcompression testing of hard materials, the reader is directed to reference~\cite{19}). These tests seek to address two central questions: (i) In order to establish a baseline with which to compare the effect of alloying, what is the yield stress of pure MoSi$_2$ in the brittle [0~0~1]-direction at room temperature? (ii) How does tantalum affect the active slip systems and their critical resolved shear stresses, given that little is known about its effect?

\section{Background: Structure and Deformation of \texorpdfstring{MoSi$_2$}{MoSi2}}
MoSi$_2$ has a C11$_b$ (tetragonal BCC) structure (Strukturbericht designation) with the I4/mmm space group (No. 139) and the lattice parameters a = 0.32 nm and c = 0.786 nm~\cite{20}. The unit cell is ordered along the [0~0~1] direction in an alternating stacking sequence of one molybdenum and two silicon planes, in which every atom of both types is surrounded by ten nearest neighbours. One characteristic of this structure is the stacking of \{1~1~0\} planes in an ABAB order. MoSi$_2$ has a different structure above 1900$^{\circ}$C, which is a hexagonal modification (C40-structure) with the space group P6$_2$22 and a stacking of (0~0~0~1) planes in the sequence ABCABC.

Umakoshi et al.~\cite{21} were the first to systematically study the plastic deformation behaviour and operating slip systems in MoSi$_2$. They determined that MoSi$_2$ single crystals could only be plastically deformed above 1000$^{\circ}$C, independent of their orientation. Furthermore, the obtained strength was highly dependent on the orientation, being highest close to [0~0~1]. They attributed the high anisotropy to the activation of two different slip systems: only the hard (high CRSS) \{0~1~3\}$<$3~3~1$>$ slip system was activated in orientations close to [0~0~1], whereas in orientations away from [0~0~1], deformation on the soft (low CRSS) \{1~1~0\}$<$3~3~1$>$ slip system could also be detected. The study published by Kimura et al.~\cite{22} is in good agreement with the results of Umakoshi et al.~\cite{21}. The works of Mitchel and Maloy~\cite{23} and Maloy et al.~\cite{24} are controversial concerning the possible slip systems in MoSi$_2$, identifying the operating slip systems as \{0~1~3\}$<$3~3~1$>$ below 1300$^{\circ}$C and \{0~1~1\}$<$1~1~1$>$ above 1300$^{\circ}$C in the [0~0~1] orientation, and \{1~1~0\}$<$1~1~1$>$ and \{0~1~1\}$<$1~0~0$>$ in all other deformation directions. Nevertheless, all agree concerning the temperature at which plastic flow becomes feasible (1000$^{\circ}$C), and all show a strong orientation dependence of the yield strength.

\begin{figure}[ht]
	\centering
	\includegraphics[width=.6\linewidth]{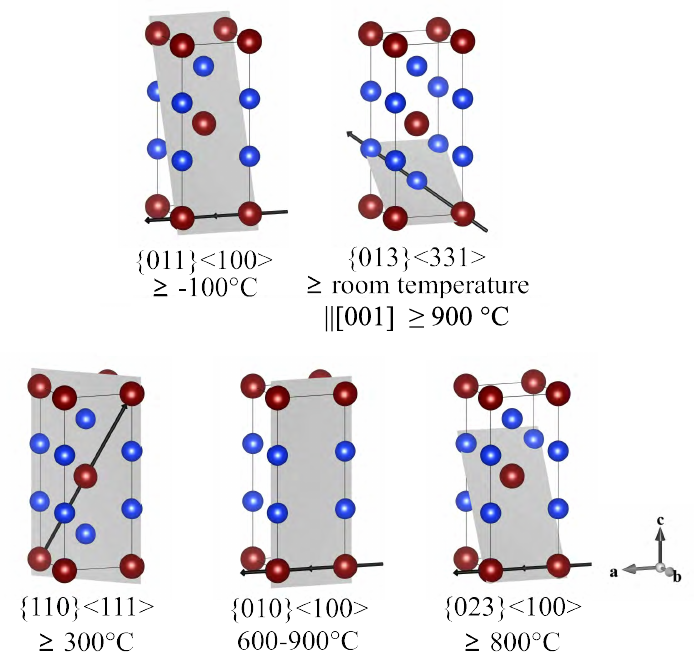}
	\caption{Slip systems in MoSi$_2$ (grey) and their active temperature regime, from references~\cite{1,2}. Si atoms are coloured red, Mo atoms blue.} 
	\label{Figure1}
\end{figure}

According to Ito et al.~\cite{1,2,25} MoSi$_2$ will plastically deform at room temperature but only in orientations away from [0~0~1]. Single crystals in an [0~0~1] orientation could only be deformed above 900$^{\circ}$C. Ito et al. performed single crystal compression experiments in three different orientations ([0~15~1], [1~1~0] and [2~2~1]) at varying temperatures, with the activated slip systems determined by subsequent TEM studies. The five different slip systems that were identified and their active temperature regime are illustrated in Figure~\ref{Figure1}. They observed dislocation motion on \{0~1~1\}$<$1~0~0$>$ and \{0~1~3\}$<$3~3~1$>$ slip systems at room temperature, whereas deformation on the other slip systems only took place at temperatures higher than 300$^{\circ}$C~\cite{1}. In general, slip on \{0~1~1\}$<$1~0~0$>$ followed Schmid’s law, but in contrast, studies on \{0~1~3\}$<$3~3~1$>$ show the highest CRSS in the exact [0~0~1] orientation and a CRRS value close to the other known slip systems in orientations away from [0~0~1]~\cite{1,2,21,22,23,24}. That is, Schmid’s law is not valid for slip on \{0~1~3\}$<$3~3~1$>$ slip systems, because the CRSS strongly depends on crystal orientation. Ito et al.~\cite{2} related this phenomenon to the dependence of the various different dissociation and decomposition reactions on crystal orientation, dislocation line orientation and temperature; sessile configurations can occur for three dislocation line orientations in specimens deformed near [0~0~1], but only for two line orientations away from [0~0~1]. More recently, Paidar et al.~\cite{26,27} have considered the dislocation core structure and energy of the associated stacking faults in the movement of partial dislocations: the existence of asymmetric faults as well as non-planar dissociations suggests that an orientation dependence of the critical resolved shear stress is to be expected, with the $\frac{1}{2}<$3~3~1$>$ dislocations in C11$_b$ behaving analogously to $\frac{1}{2}<$1~1~1$>$ BCC dislocations.

Plastic deformation in polycrystalline MoSi$_2$ takes place on four independent slip systems, which are the most frequently observed \{0~1~1\}$<$1~0~0$>$ and \{1~1~0\}$<$1~1~1$>$ slip systems and all combinations of \{0~1~1\}$<$1~0~0$>$, \{1~1~0\}$<$1~1~1$>$, \{0~1~0\}$<$1~0~0$>$ and \{0~2~3\}$<$1~0~0$>$~\cite{28}. In order to achieve an arbitrary shape change in a polycrystal by slip (in the absence of sufficiently fast diffusional mechanisms at room temperature), the von Mises criterion of five independent slip systems must be fulfilled~\cite{29} and slip therefore must also occur on \{0~1~3\}$<$3~3~1$>$. This slip system is therefore key for polycrystalline ductility.

To enhance the room temperature ductility and fracture toughness of polycrystals, the high orientation dependence and critical resolved shear stress of \{0 1 3\}$<$3 3 1$>$ slip has to be minimised, or another independent slip system has to be activated in addition to those observed during room temperature deformation. This can be realised by a decrease of the stacking fault energy on distinct slip planes. For this purpose transition metals (TM) are suitable microalloying elements that form disilicides with the tetragonal C11$_b$, the hexagonal C40 or the orthorhombic C54 structure. All of these structures show hexagonally-ordered TMSi$_2$ planes but their stacking sequence differs: C11$_b$: AB, C40: ABC, and C54: ADBC~\cite{11}. Studies regarding the solubility of ternary alloying elements in MoSi$_2$ showed that these elements can be divided into three groups~\cite{12}:
\begin{itemize}
	\item W and Re substitute molybdenum and form a solid solution with MoSi$_2$.
	\item Cr, Zr, Nb, Ta and Al are soluble in MoSi$_2$ in low concentrations; at higher concentrations a second C40-phase forms and builds a lamellar microstructure or coarse precipitates. Cr, Zr, Nb and Ta substitute Mo-atoms, Al substitutes Si-atoms.
	\item  Additions of Ru and Hf do not dissolve in MoSi$_2$ and therefore lead to the formation of second phases. This can lead to a lamellar-like two-phase microstructure if the contents of alloying elements are high enough and the phases solidify in a eutectic reaction. 
\end{itemize}

The alloying elements Al, Nb and Ta studied in this work have limited solubilities in MoSi$_2$ which are x = 0.045 in Mo(Si$_{(1-x)}$Al$_x$)$_2$), x = 0.039 in (Mo$_{(1-x)}$Nb$_x$)Si$_2$ and x = 0.150 in (Mo$_{(1-x)}$Ta$_x$)Si$_2$ respectively. Exceeding these values causes the hexagonal C40 phase to form with a high concentration of the alloying element~\cite{3,12,30}. Remaining in the solid-solution region, a decrease in hardness of alloyed MoSi$_2$ has been observed for all alloying elements compared to the pure material~\cite{12}.

While the \{1 1 0\}$<$1 1 1$>$ slip systems can only be activated above 300$^{\circ}$C in pure MoSi$_2$ (see Figure~\ref{Figure1}), they have been observed at room temperature in microalloyed MoSi$_2$~\cite{11}. According to Peralta et al.~\cite{14}, microalloying with Al lowers the critical shear stress of this particular slip system. The $\frac{1}{2}<$1 1 1$>$ dislocations dissociate into two identical $\frac{1}{4}<$1 1 1$>$ partial dislocations, separated by a stacking fault on \{1~1~0\}~\cite{1,28,31}. The stacking of planes in the disordered stacking fault region is ABC which is analogous to the order of (0~0~0~1) planes in the hexagonal C40 structure. The microalloying with C40-stabilising elements leads to a decrease of the stacking fault energy between $\frac{1}{4}<$1~1~1$>$ partial dislocations which affects the Peierls stress and the dislocation mobility~\cite{14}. Consistent with this expectation, Peralta et al.~\cite{14} observed a wider dissociation width of partial dislocations in MoSi$_2$ containing Al or Nb. A decrease of the stacking fault energy was also found via theoretical examinations by Waghmare et al.~\cite{8,13}.

For the sake of completeness it should be mentioned that the different alloying elements, like Al or Nb, showed a softening effect at room temperature but also solid solution hardening at very high temperatures~\cite{32}. The underlying mechanisms are not fully understood. The appearance of yield stress anomalies on several slip systems is evidence that the mechanisms observed at room temperature do not extend in a constant fashion towards elevated temperatures and a complete understanding throughout all temperature regimes is needed. However, until now, the major experimental challenge was encountered in the brittle, low temperatures regime and therefore not all yield or critical resolved shear stresses essential for a purposeful improvement of ductility in MoSi$_2$ and its alloys have been accessible.

\section{Experimental Procedure}
\subsection{Materials and Sample Preparation}
Polycrystalline discs were produced by arc-melting pure molybdenum and silicon in the stoichiometric ratio of 1:2 with several remelting cycles to ensure homogeneity. This alloy served as the reference material. Additionally, three discs were microalloyed ($<$3 at\%) with niobium, tantalum or aluminium. The arc-melted discs were cut into parallel slices, ground and polished with a chemo-mechanical OPS finish. EDX (energy dispersive X-ray spectroscopy) and EBSD (electron backscatter diffraction) measurements (Zeiss 1540 EsB with EDX- and EBSD-system by Oxford Instruments) were performed to determine the alloying compositions (Table \ref{Table1}) and confirm the obtained crystal structures and their orientations. The concentrations of the microalloyed samples were limited by those experimentally possible with the arc melting process, rather than a solubility limit for example.

\begin{table}[ht]
\centering
\begin{tabular}{c c}
\hline
Alloy & Concentration of Microalloying Element (at\%) \\
\hline
MoSi$_2$ & - \\
Mo(Si,Al)$_2$ & 2.0 $\pm$ 0.8 \\
(Mo,Nb)Si$_2$ & 1.2 $\pm$ 0.4 \\
(Mo,Ta)Si$_2$ & 1.0 $\pm$ 0.3 \\
\hline
\end{tabular}
\caption{Concentration of microalloying elements as measured by EDX \label{Table1}}
\end{table}

\subsection{Nanoindentation}
Nanoindentation measurements on polycrystalline MoSi$_2$, Mo(Si,Al)$_2$, (Mo,Nb)Si$_2$ and (Mo,Ta)Si$_2$ were performed a Nanoindenter G200 (Agilent Technologies) in several differently-oriented grains using a Berkovich diamond tip and the continuous stiffness method (CSM, 45~Hz, 2~nm amplitude) with maximum indentation depths of 1500~nm. Tip shape and frame stiffness were calibrated on fused silica~\cite{33} and the Poisson’s ratio MoSi$_2$ was taken to be 0.157~\cite{34}. The indents were imaged in a scanning electron microscope (SEM, Zeiss 1540EsB) and the occurring slip lines on the surface were graphically correlated to the local grain orientation measured by EBSD using the software VESTA~\cite{35}. For this analysis, about 15 indentations each with 1000~nm and 1500~nm depth were used as slip lines were most prominent at these depths. To prevent an influence of orientation on hardness and the slip line analyses, the indentations were performed in several different orientations as listed in Table \ref{Table2}.

\begin{figure}
	\centering
	\includegraphics[width=.6\linewidth]{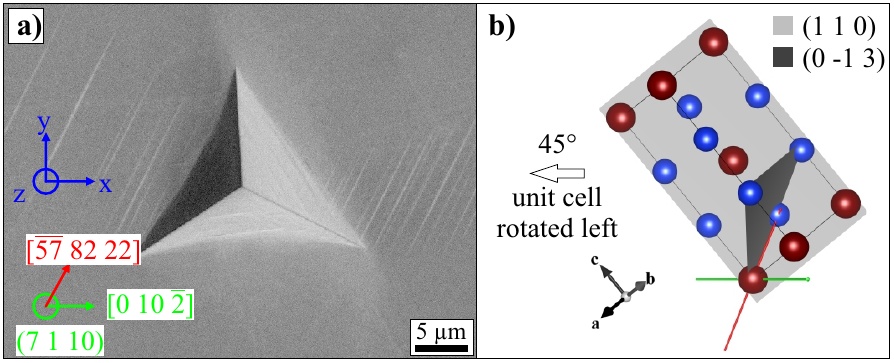}
	\caption{(a) Nanoindentation in (Mo,Nb)Si$_2$ with slip lines and (b) corresponding unit cell with 45$^{\circ}$ rotation for a better visualisation.}
	\label{Figure2}
\end{figure}

An imprint and the corresponding unit cell are illustrated in Figure \ref{Figure2} to describe the analysis of slip lines. The orientation of the unit cell (Figure \ref{Figure2}b) is parallel to the image plane and the specimen surface, i.e.~the (7~1~10) lattice plane normal is parallel to the z-axis of the coordinate system and the drawn green vector [0~10~$\overline{2}$] lies parallel to the x-axis (blue in Figure \ref{Figure2}a). The slip planes are determined via graphical analysis and correlation of the lines of intersection between the slip planes themselves and the surface.

\subsection{Transmission Electron Microscopy}
TEM membranes of indents with 500~nm indentation depths in MoSi$_2$ and (Mo,Ta)Si$_2$ were prepared by FIB (focused ion beam) milling and in-situ lift-out (Helios NanoLab, FEI) using an acceleration voltage of 30~kV with a final milling current of 80~pA to 100~pA and imaged at 200~kV accelerating voltage in a CM200 (Phillips) to determine the activated Burgers vectors using the $g \cdot b = 0$ invisibility criterion.

\subsection{Micropillar Compression}
\begin{figure}
	\centering
	\includegraphics[width=.6\linewidth]{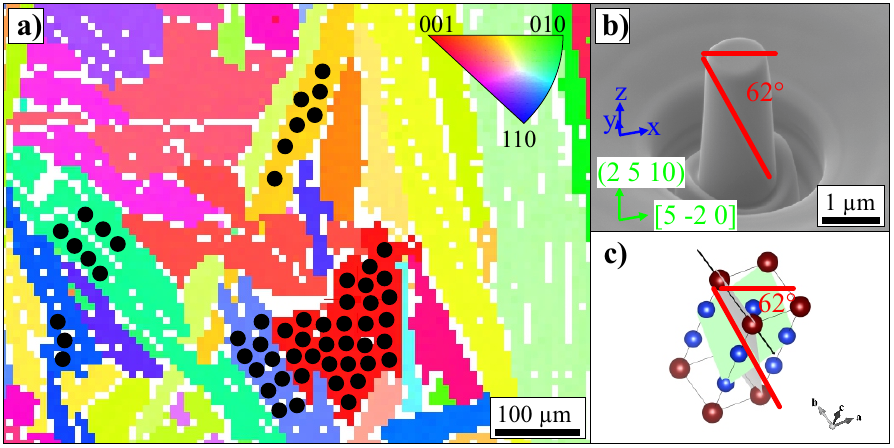}
	\caption{(a) EBSD map to determine the orientation of the micropillars with schematically drawn FIB-milled micropillars (black circles). (b) SEM image of a compressed micropillar in MoSi$_2$ in lateral view with single slip deformation. The grain’s coordinate system is illustrated in blue, the results of EBSD measurements in green and the red line highlights the slip lines and their angle to the horizontal. (c) The unit cell is shown with a (0~1~1)[1~0~0] slip system (grey) and the slip line according to (b). The green plane in (c) represents the image plane.}
	\label{Figure3}
\end{figure}

Micropillars were fabricated in MoSi$_2$ and (Mo,Ta)Si$_2$ using a FIB (Helios NanoLab, FEI) operated at 30 kV with a final milling current of 80 pA. Based on prior EBSD measurements, grain orientations of particular interest were chosen; those were [0~0~1] (hard mode) and orientations differing from [0~0~1] (soft mode) containing orientations chosen to achieve the highest Schmid factors for the \{1 1 0\}$<$1 1 1$>$ slip system. A schematic of micropillars prepared within one EBSD map is given in Figure \ref{Figure3}. All micropillars had a circular cross-section with a top diameter of approximately 1 $\mu$m and an aspect ratio of 2:1. The taper was 4–5$^{\circ}$ on average which is a common artefact of ion milling preparation perpendicular to the surface.

The compression experiments were carried out using the Nanoindenter G200 and a diamond flat punch indenter in load-control mode to a compression depth of 120~nm to 170~nm depending on the mechanical response of the given orientation. Prior to testing, the sample surface was aligned perpendicular to the flat punch using test impressions and a specially-designed tripod sample holder~\cite{36} was used. To avoid strain rate effects, the loading rate was held constant at 500~$\mu$N/s with pillars manufactured with the same diameter. The top diameter of the pillars was used to calculate the yield stress as this was the highest-stressed region, and the displacement values were corrected after compression according to a simple Sneddon model by Frick et al.~\cite{37}, using 431~GPa and 0.157 for the Young's modulus and Poisson’s ratio of MoSi$_2$~\cite{34}.

The determination of the activated slip system was carried out using VESTA to correlate the slip lines on the micropillar and the crystal orientation measured by EBSD~\cite{35}, using SEM images taken in top and lateral view. Figure \ref{Figure3}b shows an example of a micropillar deformed in an orientation away from [0~0~1] in lateral view. Pillars were rotated until the slip lines were seen edge on, and the angles measured taking into account the SEM stage tilt and rotation. The corresponding slip systems were determined via comparison with possible slip systems, illustrated in Figure \ref{Figure3}c. It should be noted that the angle of the slip trace was not measured directly between the glide plane indicated in the unit cell and the horizontal of the image, but rather between the tilt-corrected tangential to the very front of the slip trace cutting through the cylindrical pillar volume and the plane defined by the pillar axis and the horizontal of the image. Furthermore, the slip directions were determined using the unit cell and SEM images from the top view, analogous to the analysis of slip planes. Critical resolved shear stresses were calculated from the applied stress at the onset of plastic deformation via the identified slip systems and the compression axis.

For MoSi$_2$ 17 micropillars were tested in the hard direction parallel to [0~0~1] and 51 micropillars in six orientations differing from [0~0~1]. Out of these, 11 micropillars in the [0~0~1] orientation and 20 micropillars in orientations different than [0~0~1] were suited for further analysis of the active slip systems and correlation with the measured stresses, i.e.~neither fast fracture nor extensive multiple slip occurred in these pillars. For (Mo,Ta)Si$_2$ 11 micropillars were tested in the hard orientation and 55 micropillars in soft orientations, with 7 hard micropillars and 24 soft micropillars suitable for further analysis.

\section{Results}
\subsection{Nanoindentation}
\begin{figure}[ht]
    \centering
    \begin{subfigure}[t]{0.5\textwidth}
        \centering
        \includegraphics[width=.9\linewidth]{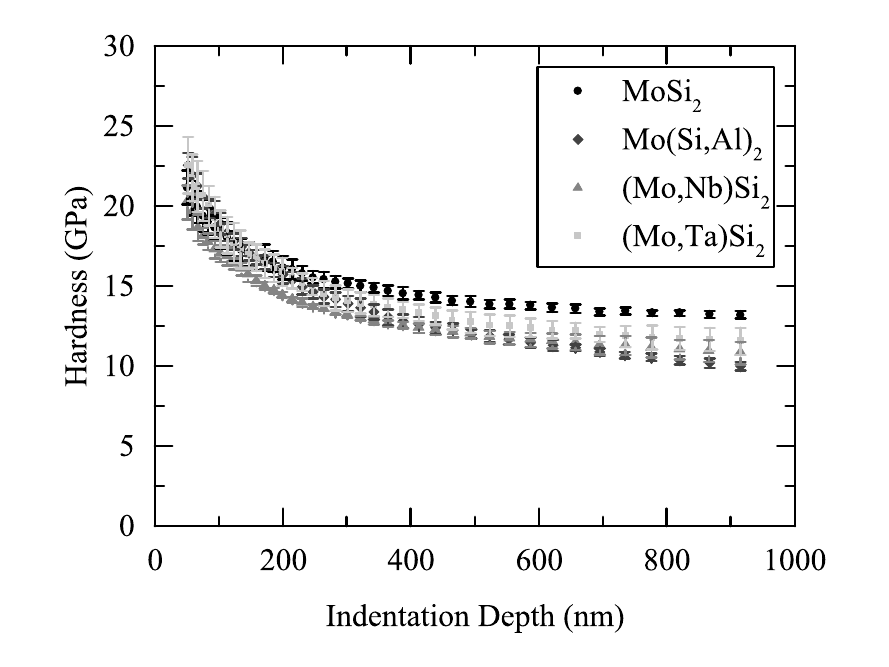}
        \caption{}\label{Figure4}
    \end{subfigure}%
    ~ 
    \begin{subfigure}[t]{0.5\textwidth}
        \centering
        \includegraphics[width=.9\linewidth]{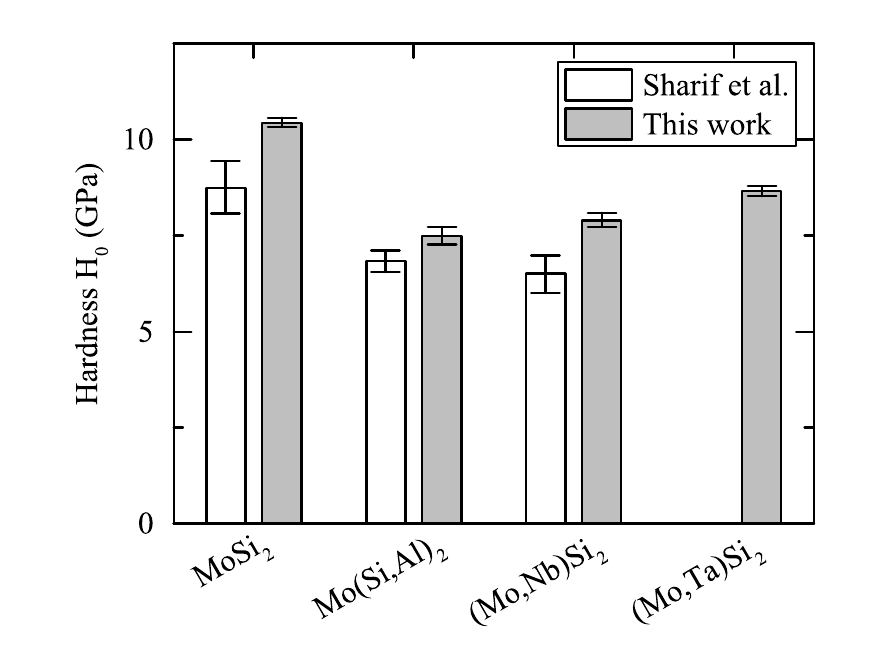}
        \caption{}\label{Figure5}
    \end{subfigure}
    \caption{(\textbf{a}) Hardness with indentation depth for MoSi$_2$, Mo(Si,Al)$_2$, (Mo,Nb)Si$_2$ and (Mo,Ta)Si$_2$. (\textbf{b}) Comparison of microindentation hardness by Sharif~\cite{32} to the extrapolated hardness from nanoindentation.}
    \label{Figure4and5}
\end{figure}

The measured hardness with indentation depth is shown in Figure~\ref{Figure4}, which exhibits a slight indentation size effect (ISE) below 500 nm. A Nix-Gao~\cite{38} fit to the data was performed to determine the extrapolated hardness $H_0$, excluding data below 150~nm due to uncertainties introduced by the CSM method at small penetration depths~\cite{39}. The corresponding decrease in hardness is highest for Mo(Si,Al)$_2$ followed by (Mo,Nb)Si$_2$, with the extrapolated values plotted in Figure~\ref{Figure5} and compared to literature~\cite{32}. Assuming a constraint factor of 2.8~\cite{40} the equivalent yield stress at 8\% plastic strain (given by the indenter geometry) has been estimated (see the Table in Figure~\ref{Table4}) to compare the nanoindentation results to micropillar compression experiments.

\subsection{Slip Line Analysis}
The slip lines around indents represent lines of intersection between a given slip plane and the surface. As the lines provide no information about the slip direction without determination of their angle with the surface, their analysis only allows an estimate of active slip planes and their activation frequency. This analysis for the four samples is given in in Table~\ref{Table2}, and shows that dislocation motion on \{0 1 3\} planes is probable for both MoSi$_2$ and microalloyed MoSi$_2$. Furthermore, dislocation motion on \{0 1 1\} planes is expected for pure MoSi$_2$. The occurrence of both slip planes is in good agreement with literature data where \{0 1 3\}$<$3 3 1$>$ and \{0 1 1\}$<$1 0 0$>$ slip systems were observed at room temperature in pure MoSi$_2$~\cite{1,2}. In all microalloyed specimens dislocation motion was found to be favoured on \{1 1 0\} instead of \{0 1 1\}.

The identification of \{0 1 0\} planes in MoSi$_2$ and Mo(Si,Al)$_2$ is most likely a result of the geometric uncertainty in the analysis (see Ref~\cite{SCHRODERS2018125} for more details). Although slip on this system was observed in previous investigations by Ito et al.~\cite{1,2}, it was restricted to the temperature regime of 600$^{\circ}$C to 900$^{\circ}$C, in broad agreement with the low frequency in this work.

\begin{table}[ht]
\centering
\begin{tabular}{c c c c c c}
\hline
Alloy & \{0 1 3\} & \{0 1 1\} & \{1 1 0\} & \{0 1 0\} & {Orientations} \\
\hline
MoSi$_2$ 		  & 22.5 & \underline{40.0} & 27.5 			   & 10.0 &  20\\
Mo(Si,Al)$_2$ & 24.6 & 14.0 			& \underline{59.6} & 1.8 & 23 \\
(Mo,Nb)Si$_2$ & 35.2 & 5.7  			& \underline{59.1} & 0.0 & 4 \\
(Mo,Ta)Si$_2$ & 39.7 & 7.4  			& \underline{52.9} & 0.0 & 13 \\
\hline
\end{tabular}
\caption{Fraction (in percent) of occurring slip planes determined via slip line analysis based on 30 indentations per alloy. Indents were performed to either 1000 or 1500 nm depth (15 each) and spread across the number of different grain orientations given. The most frequently found slip plane is underlined for each alloy. \label{Table2}}
\end{table}

\clearpage
\subsection{Transmission Electron Microscopy}
\begin{figure}
	\centering
	\includegraphics[width=.6\linewidth]{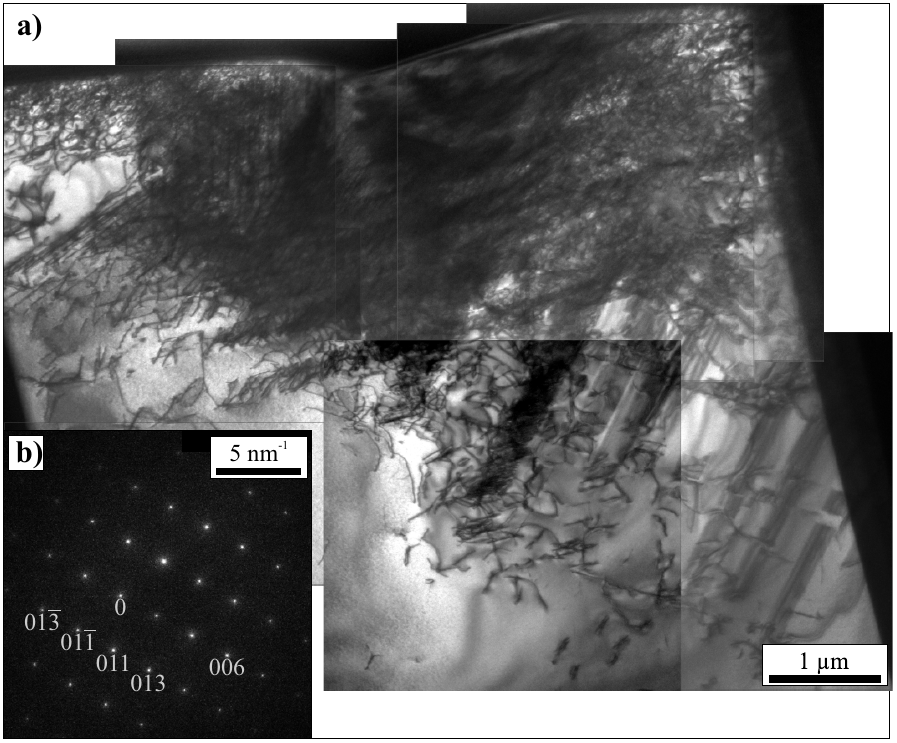}
	\caption{(a) Panorama of the plastic zone of a nanoindentation in (Mo,Ta)Si$_2$ . (b) Diffraction pattern in [1~0~0] zone axis with indexed reflexes that were analysed in two beam conditions.}
	\label{Figure6}
\end{figure}

TEM investigations were performed on MoSi$_2$ and (Mo,Ta)Si$_2$ to verify the slip systems determined via the slip line analysis. A panorama of the plastic zone of a nanoindentation in (Mo,Ta)Si$_2$ is presented in Figure~\ref{Figure6}a. The $v$ and $w$ components of Burgers vectors $[uvw]$ were determined using several two beam conditions along the [1~0~0] zone axis (Figure~\ref{Figure6}b). Therefore, in the following possible Burgers vectors are designated as $[uxx]$. 

\begin{figure}
	\centering
	\includegraphics[width=.6\linewidth]{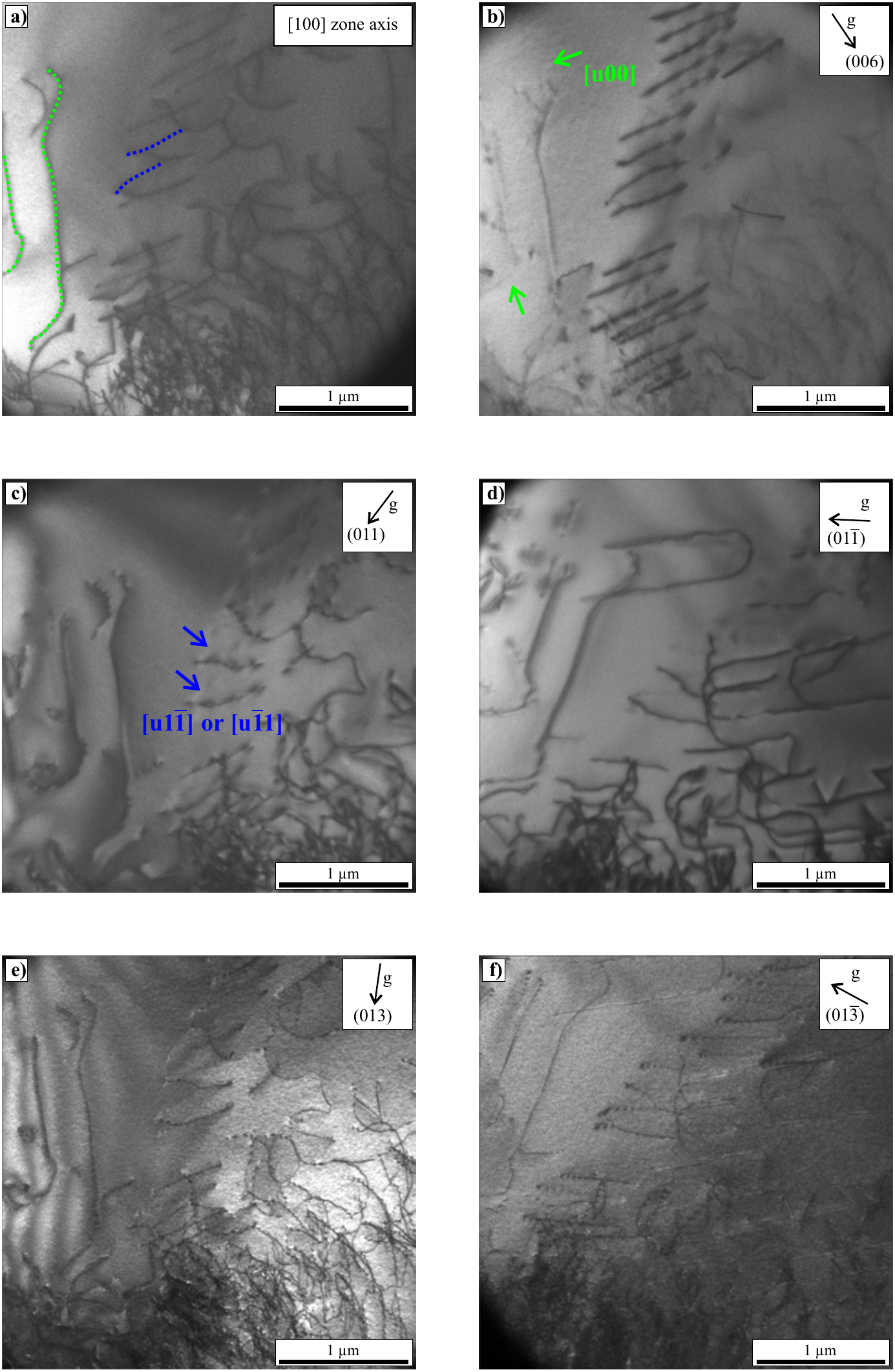}
	\caption{TEM bright field images of MoSi$_2$ (a) along the [1~0~0] zone axis and (b)-(f) in different two beam conditions. Selected dislocations and their invisibility in a certain two beam condition together with the corresponding Burgers vectors are highlighted.}
	\label{Figure7}
\end{figure}

\begin{figure}
	\centering
	\includegraphics[width=.6\linewidth]{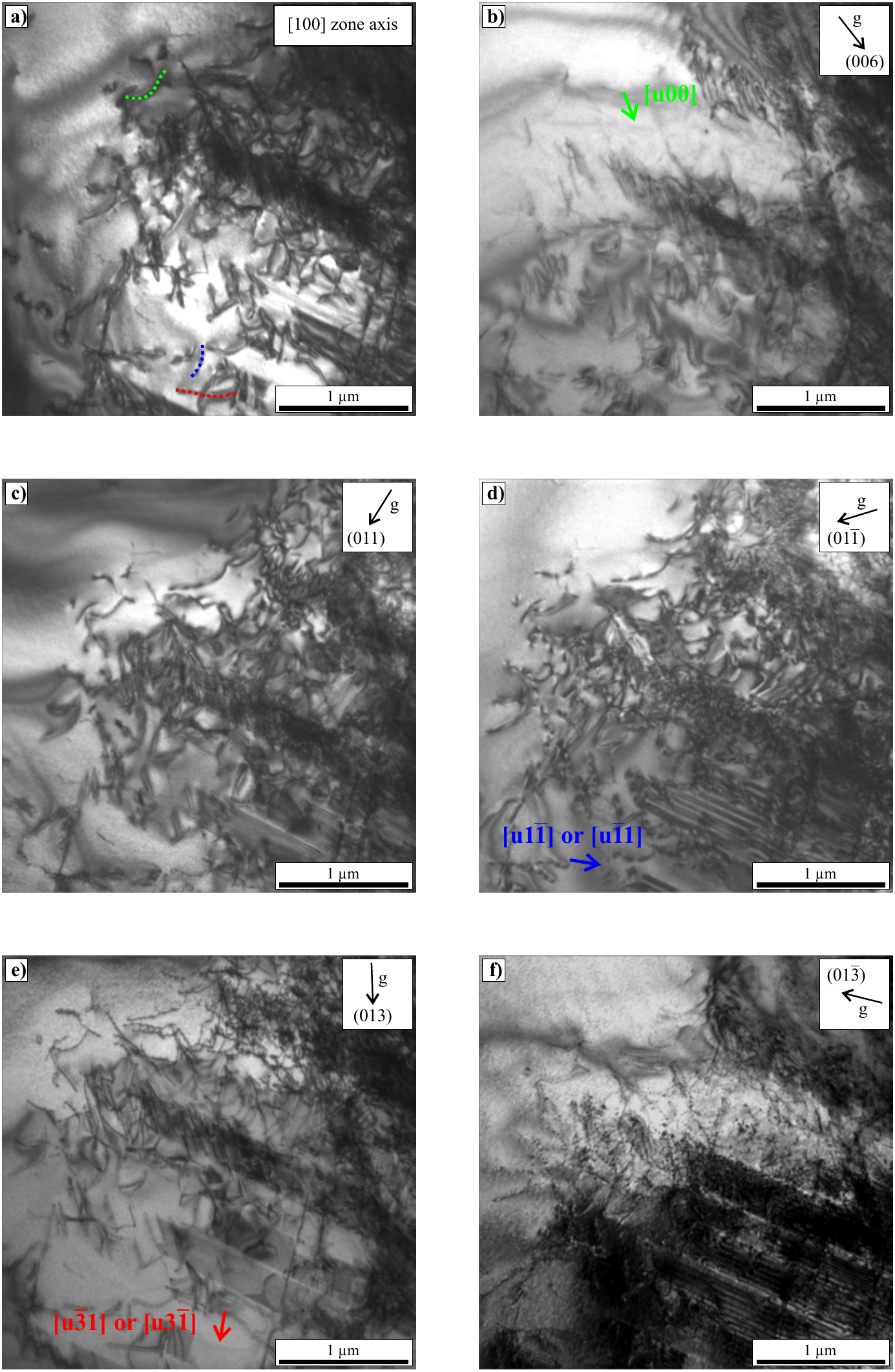}
	\caption{TEM bright field images of (Mo,Ta)Si$_2$ (a) along the [1~0~0] zone axis and (b)-(f) in different two beam conditions. Selected dislocations and their invisibility in a certain two beam condition together with the corresponding Burgers vectors are highlighted.}
	\label{Figure8}
\end{figure}

Bright field images of five two-beam conditions along the [1~0~0] zone axes of MoSi$_2$ and (Mo,Ta)Si$_2$ are presented in Figure~\ref{Figure7} and Figure~\ref{Figure8}. The images were taken on the edge of the plastic zone due to the high dislocation densities directly underneath the residual impression. The Burgers vectors determined in the chosen two beam orientations are $[uv0]$ and $[u11]$ in MoSi$_2$ and $[uv0]$, $[u11]$ and $[u31]$ in (Mo,Ta)Si$_2$. Selected dislocations are marked in Figure~\ref{Figure7} and Figure~\ref{Figure8}a) in the on-axis BF image and a diffraction condition fulfilling the invisibility criterion, i.e. $g \cdot b = 0$ with $g$ the diffraction vector and $b$ the Burgers vector. Figure~\ref{Figure8}e) and f) also show stacking faults, however the available tilting conditions precluded their identification.

\clearpage
\subsection{Micropillar Compression}
\begin{figure}[ht]
	\centering
	\includegraphics[width=.55\linewidth]{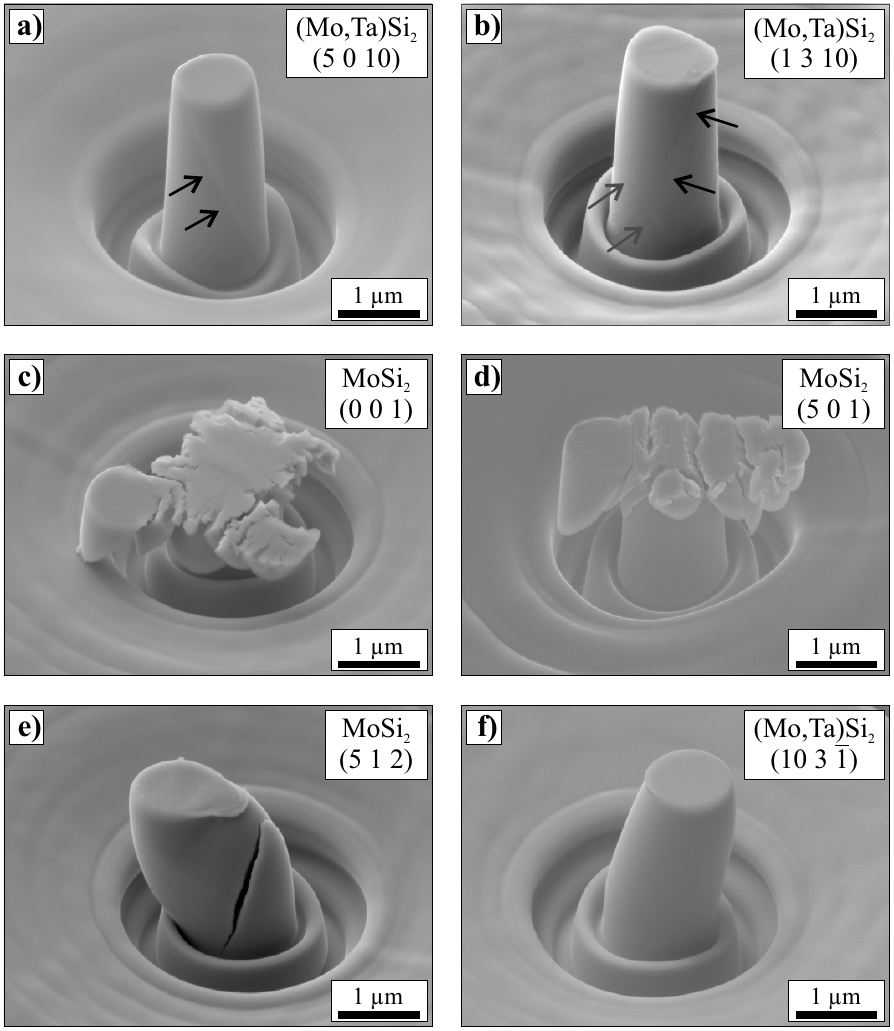}
	\caption{Different deformation mechanisms observed in compressed micropillars: (a) single slip, (b) double slip, (c) crystallographic slip with a very large strain burst, (d) splitting, (e) barrelling and cracking, (f) barrelling. Each set of parallel arrows in (a) and (b) indicates the position of a slip trace on the pillar surface.}
	\label{Figure9}
\end{figure}

The tested micropillars showed a variety of deformation mechanisms (demonstrated in Figure~\ref{Figure9}), and a complete analysis can be achieved only for those that show single slip behaviour. Pillars which slid, fractured or deformed on multiple planes can only be analysed partially. The occurrence of one particular deformation mechanism within a single orientation was repeatable and consistent. Table~\ref{Table3} summarises the compression directions, the activated slip systems and the related Schmid factors $(m)$ together with the number of tested micropillars that showed single slip.

\begin{table}
\centering
\begin{tabular}{c c c c}
\hline
Slip Systems & Compression Directions & m & No. of Tested Pillars \\
\hline
\textbf{MoSi$_2$} 		  		 & & & \\
\{013\}$<$331$>$ 				 & (001) 					 & 0.39 			& 11  \\
\{013\}$<$331$>$ 				 & (10 6 1)  				 & 0.44 			& 2   \\
\{011\}$<$100$>$				 & (5 2 10), (6 1 10)		 & 0.38, 0.29  		& 5   \\
Highest m for \{110\}$<$111$>$ & (011), (10 1 10), (3 0 5) & 0.43, 0.44, 0.46 & 13  \\
\hline
\textbf{(Mo,Ta)Si$_2$} 		 & & & \\
\{013\}$<$331$>$ 				 & (0 1 10) 				 & 0.48 			& 7  \\
\{013\}$<$331$>$ 				 & (2 1 10), (4 0 10)  		 & 0.34, 0.44		& 7  \\
\{011\}$<$100$>$				 & (5 3 4), (10 7 1)		 & 0.43, 0.43  		& 9  \\
\{110\}$<$111$>$				 & (5 0 10) 				 & 0.45				& 8  \\
\hline
\end{tabular}
\caption{Activated slip systems with their compression direction and the resulting Schmid factors m together with the number of tested micropillars. For the corresponding CRSS see text and Figure~\ref{Figure14} \label{Table3}}
\end{table}

\begin{figure}
	\centering
	\includegraphics[width=.5\linewidth]{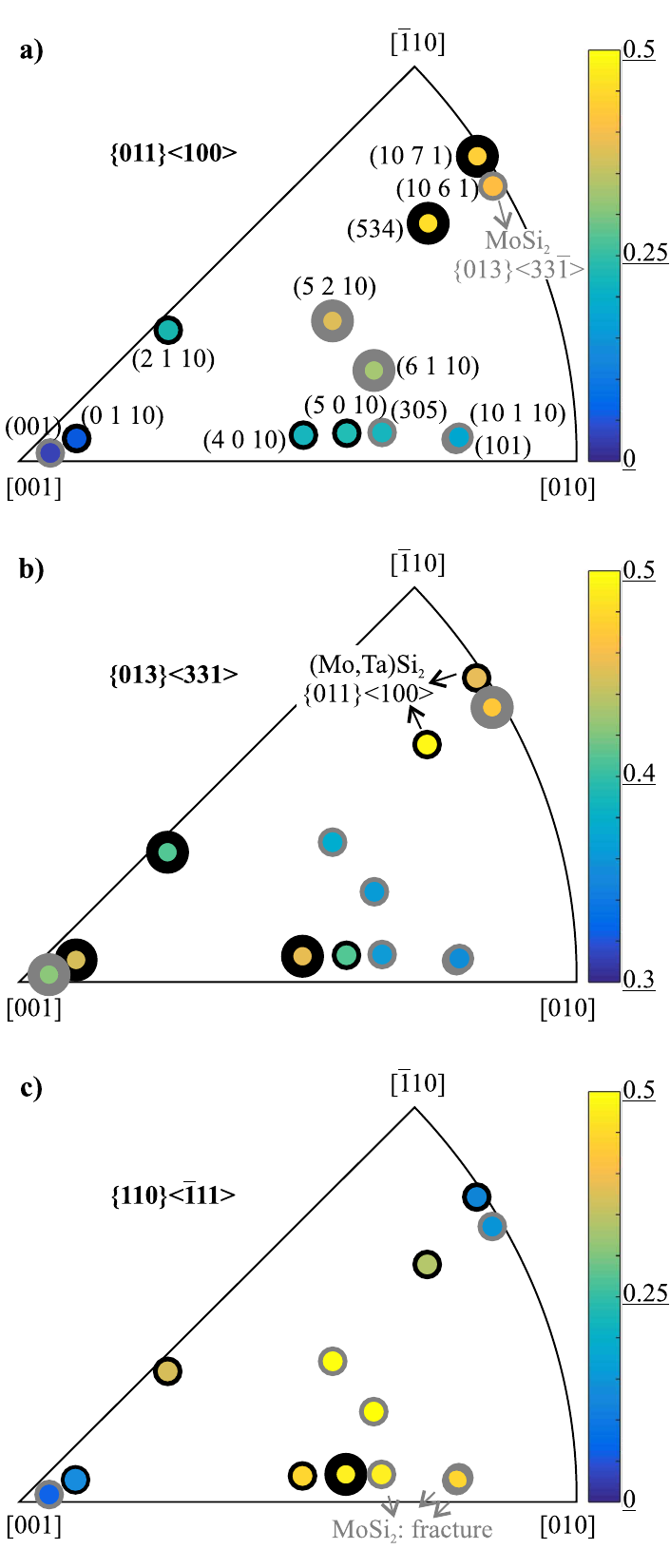}
	\caption{Inverse pole figures (IPF) giving the tested crystal orientations (see also Table \ref{Table3}) in the microcompression experiments. Each figure is colour-coded with respect to the Schmid factor on the (a) \{0 1 1\}$<$1 0 0$>$ (b) \{0 1 3\}$<$3 3 1$>$ and (c) \{1 1 0\}$<$1 1 1$>$ slip system. Orientations tested in pure MoSi$_2$ are outlined in grey, those in (Mo,Ta)Si$_2$ in black, with thicker outlines marking those samples which revealed activation of the respective slip system. Crystal directions are approximated from Euler angles with $h,k,l \leq 10$.}
	\label{Figure10}
\end{figure}

Figure~\ref{Figure10} provides colour-coded inverse pole figures for the tested crystal orientations to distinguish pure (grey ring) and alloyed (black ring) MoSi$_2$ and gives the Schmid factor on each slip system. A thicker ring indicates activation of the respective slip system and the expected correlation is largely found between a high Schmid factor and the activated slip system. Two potential exceptions are noted: (i) in Figure~\ref{Figure10}(a) and (b), the difference in the activated slip system for high Schmid factors in the vicinity of the [$\overline{1}$ 1 0] direction implies a low CRSS for the two given slip systems. (ii) the implied higher CRSS on \{1 1 0\}$<$$\overline{1}$ 1 1$>$ is in agreement with its activation only where the Schmid factor on both other systems is low and only in the alloyed (Mo,Ta)Si$_2$. It will be later shown (Figure~\ref{Figure14}) that both these implications are indeed true.

\subsubsection{Micropillar Compression Tests Close to [001]}
\begin{figure}
	\centering
	\includegraphics[width=.6\linewidth]{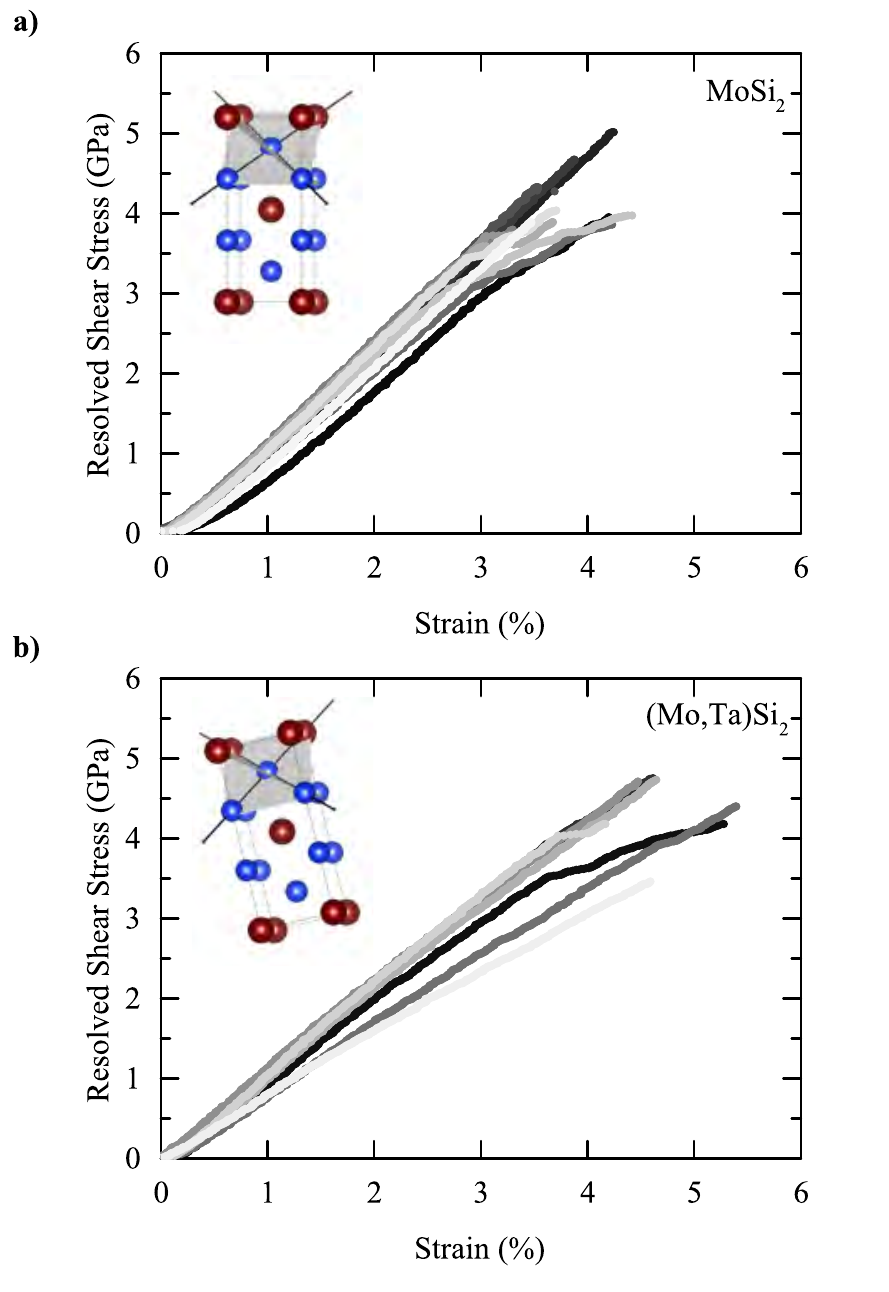}
	\caption{Resolved Shear Stresses on \{0 1 3\}$<$3 3 1$>$ slip systems versus strain for deformed micropillars (a) in [0~0~1] direction in MoSi$_2$ and (b) parallel to (0 1 10) in (Mo,Ta)Si$_2$ . The particular insets show the unit cell orientations in side view.}
	\label{Figure11}
\end{figure}

All analyses of slip planes are based on the assumption that the highest-stressed top of the micropillar plastically deformed first, i.e.~this should be the slip system on which dislocation motion was primarily activated. All micropillars close to [001], both in MoSi$_2$ and in (Mo,Ta)Si$_2$, show crystallographic slip instead of brittle fracture. For all MoSi$_2$ micropillars discussed here, all the \{0 1 3\}$<$3 3 1$>$ slip systems have an angle of 39$^{\circ}$ to the image horizontal, and therefore have the same Schmid factor of 0.39. The RSS versus strain graphs are plotted in Figure~\ref{Figure11}a, with some of the slip systems in the [0~0~1] oriented unit cell shown schematically, inset. The averaged critical resolved shear stress, determined at the first visible deviation from linear loading was 3.8 $\pm$ 0.7~GPa, rising to 4.1 $\pm$ 0.4~GPa at maximum strain.

In (Mo,Ta)Si$_2$, compression was carried out parallel to the (0~1~10) orientation, 5.7$^{\circ}$ from (0~0~1). SEM-images confirm the same plastic deformation mechanism as found in [0~0~1] oriented micropillars in MoSi$_2$ (Figure~\ref{Figure11}a), namely slip on the \{0 1 3\}$<$3 3 1$>$ slip systems, implying the small deviation from [0~0~1] does not significantly influence the deformation mechanisms. Because of this deviation, the Schmid factors of the \{0 1 3\}$<$3 3 1$>$ slip systems differ, with the highest Schmid factor of 0.48 only offered by the (1~0~3)[3~3~1] and the (1~0~3)[3~3~1] slip systems, hence the analysed micropillars only showed deformation on these two slip systems. It is suspected that slip systems with a lower Schmid factor could still be activated, which contributed to work hardening and the occurrence of a small regime of plastic flow. The plots of resolved shear stress versus strain and the (0~1~10) unit cell are given in Figure~\ref{Figure11}b). All compression experiments showed a similar stiffness and a critical resolved shear stress of 4.1 $\pm$ 0.5~GPa on \{0 1 3\}$<$3 3 1$>$ which increases to about 4.3 $\pm$ 0.3~GPa at maximum plastic strain. The CRSS for \{0 1 3\}$<$3 3 1$>$ slip may therefore actually be slightly higher in the alloyed material compared to pure MoSi$_2$.

\subsubsection{Micropillar Compression Tests in Directions Other Than [001]}
\begin{figure}
	\centering
	\includegraphics[width=.6\linewidth]{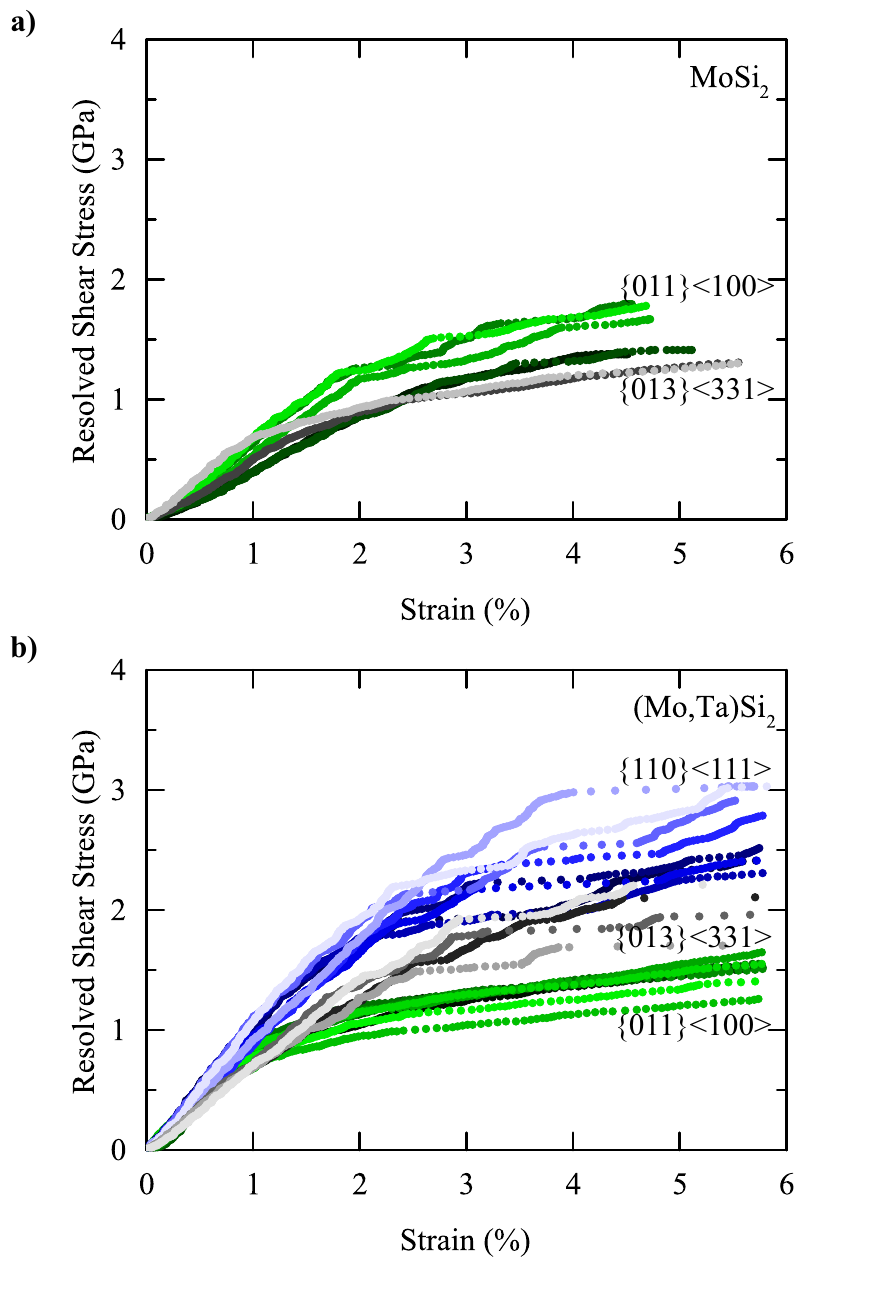}
	\caption{Resolved shear stresses versus strain for (a) MoSi$_2$ and (b) (Mo,Ta)Si$_2$ . The graphs of \{0 1 1\}$<$1 0 0$>$ slip systems are illustrated in green, \{0 1 3\}$<$3 3 1$>$ in grey and \{1 1 0\}$<$1 1 1$>$ in blue. In MoSi$_2$ slip on \{0 1 1\}$<$1 0 0$>$ and \{0 1 3\}$<$3 3 1$>$ was activated in compressions experiments parallel to (5 2 10) and (6 1 10) and (10 6 1) respectively. In (Mo,Ta)Si$_2$ slip on \{0 1 1\}$<$1 0 0$>$ was activated in compression experiments parallel to (5 3 4) and (10 7 1), slip on \{0 1 3\}$<$3 3 1$>$ parallel to (2 1 10) and (4 0 10) and slip on \{1 1 0\}$<$1 1 1$>$ parallel to (5 0 10).}
	\label{Figure12}
\end{figure}

All analysed micropillars compressed in directions other than [0~0~1] maintained a cylindrical shape and their top did not significantly slide relative to the lower end (e.g. Figure~\ref{Figure9}). MoSi$_2$ showed plastic deformation on \{0 1 1\}$<$1 0 0$>$ and \{0 1 3\}$<$3 3 1$>$ slip systems, with the respective resolved shear stresses versus strain plotted in Figure~\ref{Figure12}. Compared to the [0~0~1] orientation (Figure~\ref{Figure11}) all pillars exhibit a considerably lower stress level ($\sim$50\% lower) but show a significant regime of plastic flow. The CRSS for the \{0 1 3\}$<$3 3 1$>$ slip system is 0.7 $\pm$ 0.1~GPa parallel to (10~6~1), i.e. 85$^{\circ}$ away from [0~0~1], compared to 3.8 $\pm$ 0.7~GPa in the [0~0~1] orientation. Additional dislocation motion on \{0~1~1\}$<$1~0~0$>$ was observed in micropillars deformed parallel to (5~2~10) and (6~1~10) with an average CRSS of 1.1 $\pm$ 0.2~GPa for both orientations. Slip on \{1~1~0\}$<$1~1~1$>$ slip systems in pure MoSi$_2$ could not be observed in these orientations away from [0~0~1]. Therefore, micropillars were additionally fabricated parallel to (0~1~1), (10~1~10) and (3~0~5) to give Schmid factors of 0.46 for this particular slip system. Unfortunately, these deformed without forming clearly-assignable slip traces (Figure~\ref{Figure9}d).

Micropillars in (Mo,Ta)Si$_2$ compressed in directions other than parallel to (0~1~10) showed plastic flow instead of brittle fracture at lower stress levels compared to (0~1~10). Depending on the compression axes, dislocation motion took place on \{0 1 1\}$<$1 0 0$>$, \{0 1 3\}$<$3 3 1$>$ or \{1 1 0\}$<$1 1 1$>$ slip systems; the corresponding graphs of resolved shear stress versus strain are presented in Figure~\ref{Figure12}b). The determined critical resolved shear stresses of all slip systems are similar, with the different flow stresses  with increasing strain attributed to interaction and subsequent hardening of secondary slip systems. The \{0 1 1\}$<$1 0 0$>$ slip systems in (Mo,Ta)Si$_2$ were activated parallel to (5~3~4) and (10~7~1) and the CRSS was determined to be 0.9 $\pm$ 0.1~GPa.

Furthermore, dislocation motion on \{0 1 3\}$<$3 3 1$>$ slip systems in (Mo,Ta)Si$_2$ was observed in experiments with a compression direction parallel to the (2~1~10) and (4~0~10) plane normals, with an average CRSS of 1.1 $\pm$ 0.2~GPa. These planes have an angle between the (0~0~1) direction of 13$^{\circ}$ and 22$^{\circ}$, respectively. As it is known that Schmid's law does not apply to this slip system, these different inter-plane angles allow the orientation-dependence of this slip system to be determined. It is determined that the CRSS drops from 4.1~GPa for compression parallel to the (0~1~10) plane normal to 1.2 $\pm$ 0.2~GPa (13$^{\circ}$) and 0.9 $\pm$ 0.1~GPa (22$^{\circ}$) with an increasing angle away from [0~0~1].

In contrast to MoSi$_2$, plastic flow was observed on \{1 1 0\}$<$1 1 1$>$ slip systems in (Mo,Ta)Si$_2$ parallel to (5~0~10). (Note this direction is strictly identical to (1~0~2), but is left in this form for comparison with the compression tests in MoSi$_2$ along (5~2~10)). In this orientation the Schmid factors were highest for this particular slip system, analogous to the experiments on MoSi$_2$ as described above that resulted in brittle failure. The determined CRSS for the \{1 1 0\}$<$1 1 1$>$ slip system in (Mo,Ta)Si$_2$ is 1.3 $\pm$ 0.3~GPa.

\section{Discussion}
\subsection{Nanoindentation}
A decrease in hardness was observed for microalloyed MoSi$_2$ in comparison to the unalloyed sample. The appearance of a slight ISE in the investigated alloys is consistent with studies on other hard materials using nanoindentation but also to micropillar compression experiments in different material classes with increasing bulk strength~\cite{17,41}. The extrapolated hardness values, $H_0$, are slightly higher but still close to the microindentation results of Sharif et al.~\cite{32} (Figure~\ref{Figure5}) on pure MoSi$_2$ and the microalloyed Mo(Si,Al)$_2$ and (Mo,Nb)Si$_2$. 

The activation of the additional slip system \{1 1 0\}$<$1 1 1$>$ for plastic deformation, seen via the slip line analysis, has been described at temperatures above 300$^{\circ}$C (Figure~\ref{Figure1}) for pure MoSi$_2$, whereas the change of slip systems at room temperature in the microalloyed specimen Mo(Si,Al)$_2$, (Mo,Nb)Si$_2$ and (Mo,Ta)Si$_2$ was also specified by Inui et al.~\cite{42} and attributed to a reduced stacking fault energy. The latter has been observed experimentally in terms of an increased dissociation width of partial dislocations in MoSi$_2$ containing Al or Nb ~\cite{14} and predicted theoretically ~\cite{8,13}. The mechanism behind this may be complex, and likely affected by the interlinking effects of site occupancy (here, Al substituting Si sites, Nb and Ta substituting Mo sites), the resultant change in lattice parameter, as well as changes in electronic density and therefore bond strength and stacking fault energy and, ultimately, the activation of (new) slip systems. In complex intermetallic crystals, the interplay of these effects when adding alloying elements is not necessarily straightforward to anticipate. For example, Chen~\cite{ISI:000072734400020} showed that in the TiCr$_2$ C36 Laves phase both excess Ti and the addition of V increase the lattice parameter but give an opposite effect on hardness. The ability to conduct efficient experimental campaings using small polycrystalline samples and local plastic deformation by nanoindentation with orientation-dependent slip trace analysis may therefore be a promising approach to inform combined experimental-computational studies to further investigate the effects of alloying elements in solid solutions of MoSi$_2$ and enable the identification of new alloying additions using high-throughput computations, as applied successfully in metals ~\cite{sandlobes2017rare}.

\subsection{Transmission Electron Microscopy}
Dislocations with $[u00]$ - or more precisely [1~0~0] - as well as $[u11]$ Burgers vectors have been observed in MoSi$_2$. Slip in the plastic zone of a nanoindentation on the \{1 1 0\}$<$1 0 0$>$ slip system was also reported by Boldt et al.~\cite{43}, who additionally described slip on the \{1 1 0\}$<$0 0 1$>$ slip system. According to Boldt et al., only 12\% of the plastic zone exhibits deformation on \{1 1 0\}$<$0 0 1$>$ directly underneath the nanoindentations, which is a region that was not studied here. Therefore, the results of this work on the edge of a nanoindentation agree well with the findings and model of Boldt et al~\cite{43}. A comparison with the known slip systems in MoSi$_2$ yields $<$1 1 1$>$ as potential Burgers vectors for $[u11]$ type dislocations. However, according to Ito et al.~\cite{1} these dislocations were not expected to be activated below 300$^{\circ}$C, although the temperature regime may be expected to expand towards low temperatures due to the high stresses and stress gradients underneath a Berkovich indenter in comparison with uniaxial compression.

Burgers vectors of $[u31]$ could not be detected at the edge of the nanoindentation in MoSi$_2$. In contrast, Boldt et al.~\cite{44} described the existence of these dislocations at the edge of the plastic zone of a residual impression, but did not confirm their occurrence in later studies~\cite{45}. The activation of \{0 1 3\}$<$3 3 1$>$ slip systems was observed by other authors in compression experiments in the [0~0~1] direction at higher temperatures~\cite{21,24} and in directions away from [0~0~1] at room temperature~\cite{1}. As the corresponding \{0~1~3\} slip planes were indeed found here via both slip line analyses of nanoindentations and in micropillar compression experiments, we assume that these dislocations were simply not present in the investigated region of this particular plastic zone.

The analysis of the slip lines of nanoindentations in (Mo,Ta)Si$_2$ shows primary deformation on \{0~1~3\} and \{1~1~0\} planes, which can be correlated to the described \{0~1~3\}$<$3~3~1$>$ and \{1 1 0\}$<$1 1 1$>$ slip systems. This is consistent with the occurring Burgers vectors $[u31]$ and $[u11]$, respectively. Furthermore, the activation of these slip systems is consistent with prior investigations on Mo(Si,Al)$_2$ and (Mo,Nb)Si$_2$ by Inui et al.~\cite{11}, Peralta et al.~\cite{14} and Waghmare et al.~\cite{8,13}. They described a change of slip systems from \{0~1~1\}$<$1~0~0$>$ to \{1~1~0\}$<$1~1~1$>$ in microalloyed MoSi$_2$ due to a reduced stacking fault energy on \{1~1~0\} planes and the thereby decreased critical resolved shear stress for the \{1~1~0\}$<$1~1~1$>$ slip system.

Burgers vectors of type $[u00]$ were not described for (Mo,Ta)Si$_2$ in previous publications. However, micropillar compression experiments on (Mo,Ta)Si$_2$ do indeed show plastic deformation along \{0~1~1\}$<$1~0~0$>$ slip systems.

\subsection{Micropillar Compression Experiments}
\subsubsection{Micropillar Compression Tests Close to [001]}
\begin{figure}
	\centering
	\includegraphics[width=.6\linewidth]{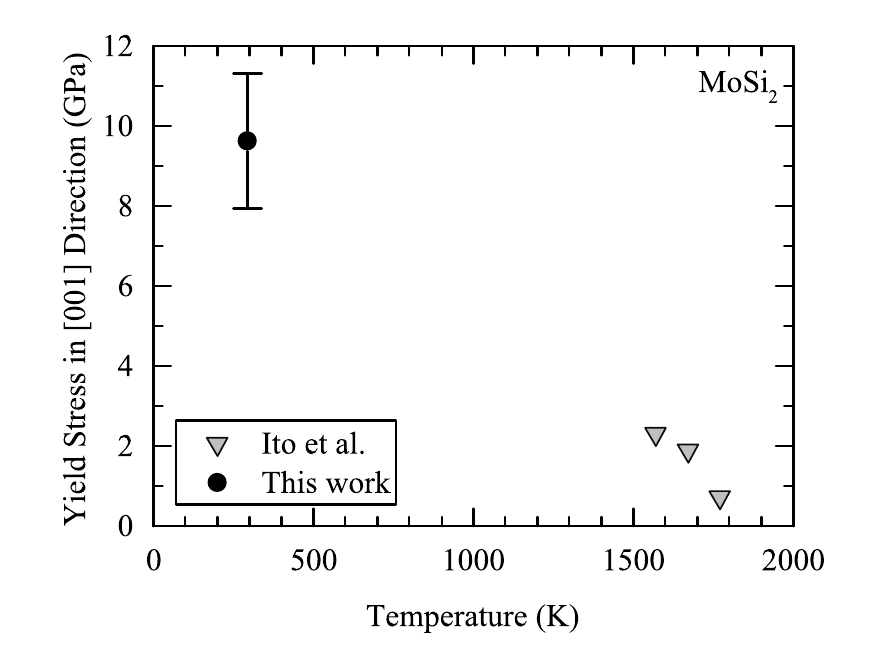}
	\caption{Comparison of compression experiments on MoSi$_2$ in [0~0~1] direction at room temperature with the results of macroscopic compression tests at high temperatures by Ito et al.~\cite{1}.}
	\label{Figure13}
\end{figure}

Micropillar compression allows a comparison between microscopic, room-temperature data and macroscopic, high-temperature data. The room temperature yield strength (at 2\% plastic flow) determined in this work is 4.2 times higher than the one at 1300$^{\circ}$C measured by Ito et al. This is consistent with the degree of thermal activation indicated by high temperature experiments (Figure~\ref{Figure13}), a more accurate analysis being limited by the lack of data points as a function of temperature from high temperature micropillar compression~\cite{46}. Assuming both that the data follow a trend of a single thermally-activated mechanism and that the CRSS is dominated by the lattice resistance to dislocation motion, the Peierls stress at 0~K can be extrapolated to give 12~GPa. This value should be taken as an upper bound estimate, given the size effects clearly present in small-scale testing (e.g.~Figure~\ref{Figure4}). 

However, in the slip systems on \{0 1 3\} with $\frac{1}{6}$ $<$3 3 1$>$ Burgers vectors~\cite{44,47}, the corresponding ratios of interplanar spacing of the slip planes and Burgers vector (d/b = 0.26 for \{0~1~3\}$<$3~3~1$>$) leads to Peierls stresses which are estimated to be between one-hundredth and one-tenth of the shear modulus~\cite{48}. Given the shear modulus of 190 GPa for MoSi$_2$~\cite{47}, the estimated Peierls stress and therefore also the measured stress at room temperature are indeed consistent with expectations. It should be noted that the relatively low nanoindentation hardness compared with these values is due to the different flow stresses of the multiple activated slip systems. 

\subsubsection{Micropillar Compression Tests in Directions Other Than [001]}
\begin{figure}
	\centering
	\includegraphics[width=.6\linewidth]{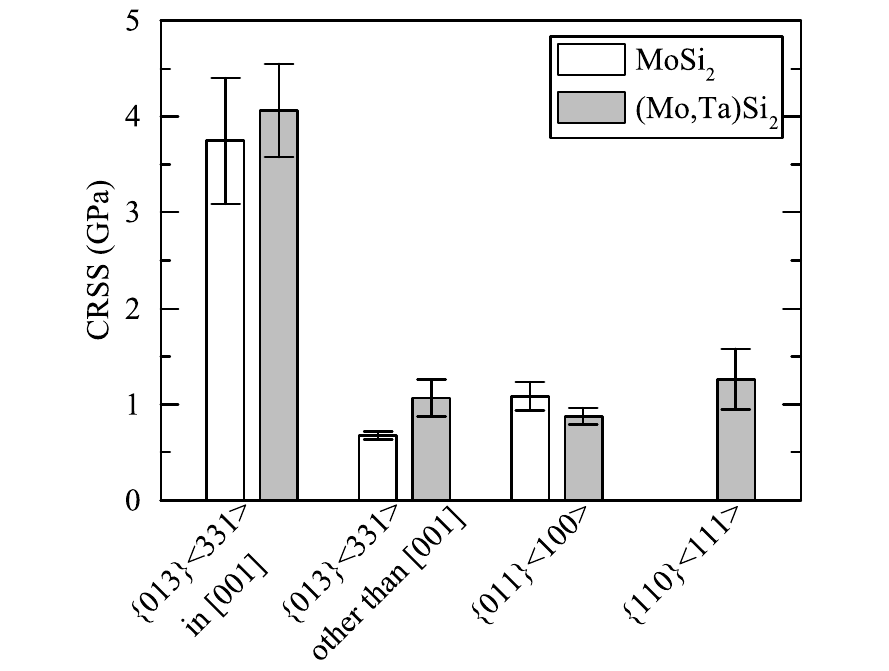}
	\caption{Critical resolved shear stresses of all observed slip systems of MoSi$_2$ and (Mo,Ta)Si$_2$.}
	\label{Figure14}
\end{figure}

Figure~\ref{Figure14} summarises all determined critical resolved shear stresses for the occurring slip systems in MoSi$_2$ and (Mo,Ta)Si$_2$. The magnitude of the stresses is in good agreement with the results of micropillar compression experiments on MoSi$_2$ single crystals performed by Nakatsuka et al.~\cite{49}. The strong orientation dependence of the CRSS of \{0 1 3\}$<$3 3 1$>$ slip, and the invalidity of Schmid's law for this slip system, were already described by Ito et al.~\cite{1} and Mitchell and Maloy~\cite{23} in MoSi$_2$. They did not observe any plastic flow and could therefore not determine the CRSS in macroscopic compression experiments along the [0~0~1] direction. However, they did observe room temperature plastic deformation along [0~15~1], i.e.~an angle of 86$^{\circ}$ to [0~0~1], which is therefore very close to the 85$^{\circ}$ of this work. Ito measured a CRSS of approximately 340~MPa, around half that measured here due to the smaller specimen volume of micropillars. These findings do however differ from microcompression experiments on MoSi$_2$ single crystals by Nakatsuka et al.~\cite{49}, who also deformed micropillars of different diameters in a [0~15~1] orientation. Their CRSS values for the \{0~1~3\}$<$3~3~1$>$ slip systems were an order of magnitude higher compared to this work, for reasons that remain unclear.

To better correlate with the macroscopic results of Ito~\cite{1}, a first estimation of the influence of specimen volume can be carried out by comparing equal plastic volumes in micropillars and indents, corresponding to an indentation depth of 100-150~nm. Since plastic deformation on all slip systems in orientations away from [0~0~1] showed similar behaviour and critical resolved shear stresses, it is assumed that the size effect is comparable for all slip systems in the softer orientations. Therefore, as the hardness at 100–150~nm is 1.5 times higher than the extrapolated macroscopic hardness, this scaling factor can be assumed for these slip systems, and is roughly the same as the difference in CRSS between the results of this work and those of Ito et al.~\cite{1}. 

Looking at the effect of tantalum, the CRSS of \{0 1 1\}$<$1 0 0$>$ slip systems in MoSi$_2$ and (Mo,Ta)Si$_2$ are very close, especially considering the standard deviation. The potentially lower CRSS in (Mo,Ta)Si$_2$ could be a result of the microalloying, corresponding to the reduced hardness of (Mo,Ta)Si$_2$ in nanoindentation. However, the fundamental mechanisms for this change have not yet been elucidated, as a decrease of stacking fault energies has so far only been described for slip systems of \{1 1 0\}$<$1 1 1$>$ type in Mo(Si,Al)$_2$ and (Mo,Nb)Si$_2$ but not for \{0 1 1\}$<$1 0 0$>$ or for (Mo,Ta)Si$_2$~\cite{8, 11, 13, 14}.

\begin{figure}
\begin{subfigure}{0.45\textwidth}
    \centering
    \includegraphics[width=1\linewidth]{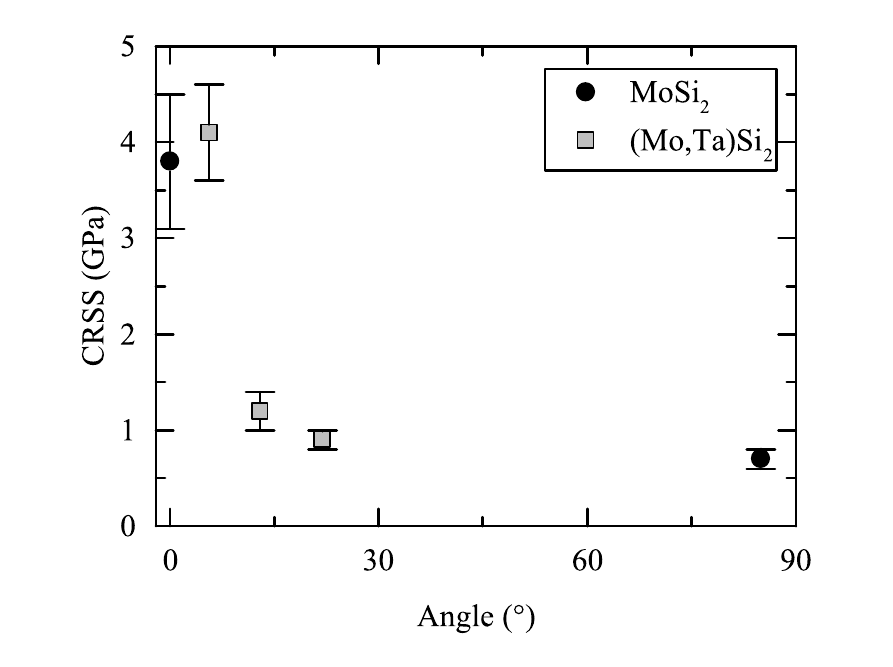}
    \caption{}\label{Figure15}
\end{subfigure}%
\begin{subfigure}{0.45\textwidth}
    \centering
\begin{tabularx}{0.8\linewidth}{c X X}
\hline
Alloy & Yield Stress at $\epsilon_{8\%}$ (GPa) & Yield Stress at $\epsilon_{5\%}$ (GPa) \\
	  & \textit{Indentation}						    & \textit{Micropillars} \\
\hline
MoSi$_2$ & $\sim$ 4.4 & 4.7 $\pm$ 0.2 \\
(Mo,Ta)Si$_2$ & $\sim$ 3.5 & 3.3 $\pm$ 0.2 \\
\hline

\end{tabularx}
    \caption{}\label{Table4}
\end{subfigure}%
\caption{(\textbf{a}) Critical Resolved Shear Stress of \{0 1 3\}$<$3 3 1$>$ slip systems versus the angle between the compression direction and the hard [0~0~1] orientation for MoSi$_2$ and (Mo,Ta)Si$_2$. (\textbf{b}) Comparison of equivalent yield stress at 8\% plastic strain from nanoindenation and average yield stress at 5\% plastic strain measured across all micropillar compression tests}
\label{Figure15andTable4}
\end{figure}

Also in (Mo,Ta)Si$_2$, the CRSS decreases with an increasing angle away from [0~0~1] to 1.2 $\pm$ 0.2 GPa or 0.9 $\pm$ 0.1 GPa, depending on the operating slip system, supporting the hypothesis of non-Schmid flow~\cite{1, 23}. This is in good agreement with the results in MoSi$_2$ where the CRSS drops to 0.7 $\pm$ 0.1 GPa at an angle of 85$^{\circ}$ to [0~0~1]. Figure~\ref{Figure15} presents the measured critical resolved shear stresses of \{0 1 3\}$<$3 3 1$>$ slip systems in MoSi$_2$ and (Mo,Ta)Si$_2$ depending on the angle to [0~0~1] and emphasises the invalidity of Schmid’s law.

The failure from very large strain bursts in deformed micropillars of MoSi$_2$ with the highest Schmid factors for slip on the \{1 1 0\}$<$1 1 1$>$ slip system is in good agreement with the above-described active temperature regimes for slip systems; dislocation motion on \{1 1 0\}$<$1 1 1$>$ is not expected below 300$^{\circ}$C in pure MoSi$_2$ and therefore a low density of mobile dislocations is present due to the high Peierls stress. Nevertheless, given the non-linearity of the stress-strain curves before the burst took place, the underlying deformation mechanisms prior to failure might be revealed in the future through dedicated experiments to lower strains.

In comparison to MoSi$_2$, (Mo,Ta)Si$_2$ showed plastic flow on \{1 1 0\}$<$1 1 1$>$ slip systems when oriented along a (5~0~10) plane normal. In this orientation the Schmid factors were highest for this particular slip system, analogous to the experiments on MoSi$_2$. The \{1 1 0\}$<$1 1 1$>$ slip systems indeed show plasticity and continuous yielding in (Mo,Ta)Si$_2$ (see Figure~\ref{Figure12}b, blue graphs), which can be attributed to a lower stacking fault energy  and therefore easier dislocation glide on \{1 1 0\} planes as in Mo(Si,Al)$_2$ and (Mo,Nb)Si$_2$~\cite{8, 11, 13, 14}. The CRSS is only slightly higher ($\sim$20\%) compared to the other slip systems observed in orientations away from [0~0~1] and therefore microalloying provides an additional slip system to improve bulk ductility. The activation of this additional slip system and lower stacking fault energy is in good agreement with the decrease in indentation hardness due to the availability of further slip systems.

Finally, comparing the averaged yield stress at 5\% plastic strain of the micropillar compression experiments with the equivalent yield stress at 8\% plastic strain determined via nanoindentation using a constraint factor of 2.8~\cite{40} the values are consistent within each material (Table~\ref{Table4}). Both determined yield stresses of (Mo,Ta)Si$_2$ are reduced compared with pure MoSi$_2$ which, as mentioned, might be attributed to the decreased stacking fault energy on \{1 1 0\} planes and particularly to the additional \{1 1 0\}$<$1 1 1$>$ slip system which enhances the ductility in the inhomogeneous, 3D stress field induced in indentation.

\section{Conclusions}
The influence of the microalloying elements aluminium, niobium and tantalum on the plastic deformation and mechanical properties of MoSi$_2$ and the underlying slip systems was studied using nanoindentation, transmission electron microscopy and micropillar compression tests. Plastic deformation could be achieved for the first time in both pure and microalloyed MoSi$_2$ by micropillar compression experiments close to [0~0~1] and in orientations away from [0~0~1] allowing the determination of (critical) resolved shear stresses.

The main conclusions of this work are:
\begin{itemize}
	\item The hardness of all microalloyed samples, i.e. Mo(Si,Al)$_2$, (Mo,Nb)Si$_2$ and (Mo,Ta)Si$_2$, is lower compared to pure MoSi$_2$.
	\item Indentation slip line analysis suggests plastic deformation on  \{0~1~3\}$<$3~3~1$>$ slip systems in pure MoSi$_2$, Mo(Si,Al)i$_2$, (Mo,Nb)Si$_2$ and (Mo,Ta)Si$_2$ and, in addition, a change in the favoured slip system from \{0~1~1\}$<$1~0~0$>$ to\{1~1~0\}$<$1~1~1$>$ in the MoSi2 alloys. 
	\item Microcompression of Ta alloyed MoSi$_2$ confirmed the activation of the additional slip system \{1~1~0\}$<$1~1~1$>$, whereas no evidence of activity on this system could be found in microcompression of pure MoSi$_2$ (note that the Burgers vector was derived from the micropillar experiments, as only slip planes could be identified from the slip line analysis).
	\item TEM studies on nanoindentations with 500~nm indentation depths in MoSi$_2$ and (Mo,Ta)Si$_2$ showed dislocations with Burgers vectors $<u00]$ and $<u11]$ and $<u00]$, $<u11]$ and $<u31]$ respectively.
	\item Tantalum as microalloying element has a negligible effect on the CRSS on the \{0~1~3\}$<$3~3~1$>$ and \{0~1~1\}$<$1~0~0$>$ slip systems.
	\item The strong orientation dependence of \{0~1~3\}$<$3~3~1$>$ slip systems described in the literature could be confirmed.
\end{itemize}

It was shown that the combination of the techniques used here are ideally suited to study the currently limited knowledge of the plasticity of brittle, high-temperature materials like MoSi$_2$ or microalloyed MoSi$_2$, where valuable statistics and average properties can be obtained by indentation, while microcompression in conjunction with SEM and particularly EBSD allows the determination of anisotropic properties and specific critical resolved shear stresses. Transmission electron microscopy on deformed samples, covering cross-sections of entirely deformed volumes, serves to confirm and study the deformation mechanisms deduced from the micromechanical experiments, such as the confirmation of particular Burgers vectors in more detail.

\section*{Acknowledgements}
The authors gratefully acknowledge financial support by the German Research Foundation (DFG) via the Research Training Group 1896 ”In Situ Microscopy with Electrons, X-Rays and Scanning Probes” and the cluster of excellence ”Engineering of Advanced Materials” at the Friedrich-Alexander-Universit\"{a}t Erlangen-Nürnberg. We also thank Dr. Konstantin Molodov (RWTH Aachen University) for his help with MTEX.

\clearpage
\section*{References}

\bibliography{mybibfile}

\end{document}